\documentclass{elsart}

\usepackage{amssymb}

\begin{document}

\begin{frontmatter}



\title{On the Range of Validity of Integral Transform Methods in Tsallis Statistical Mechanics}

\author{M. R. C. Solis} and
\ead{msolis@nip.upd.edu.ph}
\author{J. P. H. Esguerra\corauthref{cor1}}
\corauth[cor1]{Corresponding Author}
\ead{pesguerra@nip.upd.edu.ph}
\address{Theoretical Physics Group, National Institute of Physics, University of the Philippines, Diliman, Quezon City, 1101 Philippines}



\begin{abstract}
We show that if we require positive definite probabilities, then frequently cited results of
Prato, of Lenzi and Hilhorst on the nonextensive equilibrium statistical mechanics of gases are
valid only for a limited range of the Tsallis parameter, $q$. We determine the range of validity
of the Hilhorst and the Lenzi formulae. We then use various integral representations of the
Gamma function to derive new formulae for one-dimensional gases whose range of validity is
wider than that of the Hilhorst and the Prato formulae. We then apply the new formulae to the
classical ideal gas and the Tonks gas.
\end{abstract}

\end{frontmatter}

\section{Introduction}
In 1988, Tsallis postulated a new entropy function \cite{Tsallis}. He was, in part, motivated
by the difficulties of the standard thermodynamics with nonextensive systems. Since that time, his
methods have found various applications. (See references \cite{Mendes} 
to \cite{SimAnneal}.)

One of the main difficulties in using the Tsallis formalism is the proper incorporation of the
constraints. The calculation of thermodynamical quantities in this generalized scheme is more
involved than in the usual statistics, frequently making necessary the recourse to numerical
approaches\cite{Prato}.

Analytic approaches, however, have great value because because these may suggest new approximation
schemes where numerical methods might fail. An example of these analytic methods are integral
transform methods.

It has been claimed that the Tsallis partition function is an integral
transform of the corresponding q=1 partition function. Tsallis\cite{HilhorstFormula1},
Prato\cite{Prato}, and Lenzi\cite{Lenzi} have developed various integral transform formulae.

Intergral transform formulae\cite{Prato}\cite{Fa}\cite{HilhorstFormula1} have been
used in the past
to study the Tsallis statistical mechanics of the classical ideal gas\cite{Prato}, black
body radiation \cite{Mendes}, and two-dimensional self-gravitating systems\cite{Fa}, among others.
If Hilhorst-type formulae are invalid for some values of $q$, then these systems need to be
restudied for these values of $q$.

Our task is to reexamine the derivation of the integral transform formulae, and to examine
their respective domains of validity. In the following discussion, we concentrate on the canonical
ensemble. In particular, we use the unnormalized constraints formalism\cite{TsallisMendes}.

\subsection{The Hilhorst Formula}

The Tsallis partition function for the canonical ensemble is
\begin{eqnarray}
Z_q={\sum_i }(1+\beta(q-1)H_i)^{\frac{1}{1-q}}\label{TsallisZ}
\end{eqnarray}
Note that this is a {\em restricted} sum; the sum is only over all states
for which $1+\beta(q-1)H_i \ge 0$. The $H_i$'s are the eigenvalues of the Hamiltonian.

It is now convenient to introduce the Gamma function so that the power law form may be written
as an integral of an exponential. The representation of the Gamma function to be used is given by
\begin{eqnarray}
\Gamma(y)=x^y\int_0^{\infty}t^{y-1}\mbox{e}^{-xt}\mbox{dt}
\label{A}
\end{eqnarray}
$\mbox{where Re(x)}>0,\mbox{Re(y)}>0.$.

From (\ref{A}), it can easily be seen that, for $\mbox{Re(x)}>0,\mbox{Re(y)}>0.$
\begin{eqnarray}
x^{-y}&=&\frac{1}{\Gamma(y)}\int_0^{\infty}t^{y-1}\mbox{e}^{-xt}\mbox{d}t
 \label{power}
\end{eqnarray}

Setting $x=1+\beta(q-1)H_i$, and $y=\frac{1}{q-1}$, the expression for $p_i$
may be written as
\begin{eqnarray}
[1+\beta(q-1)H_i]^{\frac{1}{1-q}}=\frac{1}{\Gamma(\frac{1}{q-1})}\nonumber\\ \int_0^{\infty}
t^{\frac{2-q}{q-1}}
\mbox{e}^{-t[1+\beta(q-1)H_i]}\mbox{d}t\label{prev}
\end{eqnarray}
where $\mbox{Re}(1+\beta(q-1)H_i)>0$, $q>1$.

Previous workers then went on to substitute (\ref{prev}) into the partition function to obtain
\begin{eqnarray}
Z_q&=&{\sum_i}\frac{1}{\Gamma(\frac{1}{q-1})}\int_0^{\infty}t^{\frac{2-q}{q-1}}\mbox{e}^{-t[1+
\beta(q-1)H_i]}\mbox{d}t\\
&=&\frac{1}{\Gamma(\frac{1}{q-1})}\int_0^{\infty}t^{\frac{2-q}{q-1}}{\sum_i}
\mbox{e}^{-t[1+\beta(q-1)H_i]}\mbox{d}t\\
&=&\frac{1}{\Gamma(\frac{1}{q-1})}\int_0^{\infty}t^{\frac{2-q}{q-1}}\mbox{e}^{-t}{\sum_i}
\mbox{e}^{-t[\beta(q-1)H_i]}\mbox{d}t\label{dugas1}
\end{eqnarray}

Previous workers then went on to write
\begin{eqnarray}
Z_q=\frac{1}{\Gamma(\frac{1}{q-1})}\int_0^{\infty}t^{\frac{2-q}{q-1}}\mbox{e}^{-t}
Z_1(t(q-1)\beta)\mbox{d}t\label{dugas2}
\end{eqnarray}
This is called the Hilhorst formula.

\subsection{The Lenzi formula}
In this section, the derivation of the Lenzi formula\cite{Fa},\cite{Lenzi} is reviewed.

To obtain the Lenzi formula, consider the following representation of the Gamma
function\cite{Gradshteyn}:
\begin{eqnarray}
\frac{1}{\Gamma(z)}=\frac{\mbox{exp}(ab)b^{1-z}}{2\pi}\int_{-\infty}^{\infty}\frac{\mbox{e}^{itb}}
{(a+\mbox{i}t)^z}\mbox{d}t
\end{eqnarray}
with $a>0$,$b>0$, Re($z$)$>0$, and $-\pi/2<\mbox{arg}(a+\mbox{i}t)<\pi/2$.

Setting $a=1$, $b=1+\beta(q-1)H$, and $z=\frac{1}{(1-q)}+1$, it may be seen that
\begin{eqnarray}
[1+\beta(q-1)H_i]^{\frac{1}{1-q}}=\int_{-\infty}^{\infty} K_q(t)\mbox{e}^{(-(1+\mbox{i}t)(1-q)
\beta H)}\mbox{d}t\label{LenziFormula}
\end{eqnarray}
where
\begin{eqnarray}
K_q(t)=\frac{\Gamma[(2-q)/(1-q)]}{2\pi}\frac{\mbox{exp}(1+\mbox{i}t)}{(1+\mbox{i}t)^{(2-q)/(1-q)}}
\end{eqnarray}
From the form of the partition function (\ref{TsallisZ}), and from (\ref{LenziFormula})
the Tsallis partition function was written as
\begin{eqnarray}
Z_q&=&{\sum_i}\int_{-\infty}^{\infty} K_q(t)\mbox{e}^{(-(1+\mbox{i}t)(1-q)\beta)H}\mbox{d}t\\
&=&\int_{-\infty}^{\infty} K_q(t){\sum_i}\mbox{e}^{(-(1+\mbox{i}t)(1-q)\beta)H}
\mbox{d}t\label{Pratosfall}
\end{eqnarray}
The expression ${\sum_i}\mbox{e}^{(-(1+\mbox{i}t)(1-q)\beta)H}$ was then identified in
previous works as
\begin{eqnarray}
Z_{1}((1+\mbox{i}t)(1-q)\beta)={\sum_i}\mbox{e}^{(-(1+\mbox{i}t)(1-q)\beta)H}\label{dugas3}
\end{eqnarray}
Using (\ref{dugas3}), the $q<1$ partition function was written as
\begin{eqnarray}
Z_q=\int_{-\infty}^{\infty} K_q(t)\mbox{e}^{-t}Z_{1}((1+\mbox{i}t)(1-q)\beta)\mbox{d}t\label{dugas4}
\end{eqnarray}
This expression is called the Lenzi formula.

The Prato formula \cite{Prato} was obtained in a manner similar to the way the Lenzi formula was
obtained. The only difference is in the expression for the Gamma function. The Gamma function
Prato used was expressed as a contour integral, and it may be shown (by a suitable deformation
of the contour \cite{Whittaker}) that the Prato and Lenzi formulae are equivalent.

\section{How one can go wrong with integral transform methods}

We now reexamine the range of validity of the various integral transform methods. As we have
stated, integral transform formulae are important because they allow us, {\em when these
formulae are valid}, to express the Tsallis partition function of a given system in terms of
the $q=1$ or Boltzmann-Gibbs partition function. Unfortunately, as we now show, integral transform
methods are valid only under some restrictive conditions.

These restrictive conditions have their roots in the positive-definite probability requirement.
In deriving the Hilhorst formula, the sum
$\sum_{i}\mbox{e}^{-t\beta(q-1)H_i}\label{Z_1daw}$ was identified as the $q=1$ partition function, with $\beta$ replaced by
$t\beta(q-1)$. Only when this identification is valid may we deduce (\ref{dugas2}) from
(\ref{dugas1})

This identification will not be valid for all cases because the sum in (\ref{Z_1daw})
is a {\em restricted} sum. That is, we only include states for which $[1+\beta(q-1)H]\ge 0$.
On the other hand, for the $q=1$ partition function, we have an unrestricted sum. Thus, {\em only when the unrestricted sum is equal to the restricted sum will the integral transform formulae be valid.}

The Hilhorst formula will be valid when $q>1$ and all the eigenvalues $H_i$ of the Hamiltonian satisfy the inequality $H_i \ge -\frac{1}{\beta(q-1)}$. The set of positive definite Hamiltonians will obviously satisfy this requirement. Even if the Hamiltonian is not positive definite, there will be a range of values of $q>1$ and $\beta$ for which the Hilhorst formula will be valid. The Hilhorst formula will be valid when the greatest lower bound of the set of eigenvalues of the Hamiltonian is greater than or equal to $-\frac{1}{\beta(q-1)}$.

In the derivation of the Lenzi formula, the sum ${\sum_i}\mbox{e}^{(-(1+\mbox{i}t)(1-q)\beta)H}$
was identified as the $q=1$ partition function where $\beta$ is replaced by $(1+\mbox{i}t)(1-q)\beta$. But this sum is a restricted sum; we must sum only over all states for which
$1+\beta(q-1)H \ge 0$. Only when the restricted sum is equal to the unrestricted sum will the formulae of Prato and of Lenzi be valid.

The Lenzi and the Prato formula will obviously be valid when $q<1$ and the Hamiltonian is
negative definite. In fact, Lenzi and the Prato formulae will be valid if $q<1$ and the least
upper bound of the eigenvalues of the Hamiltonian is less than or equal to
$\frac{1}{\beta(1-q)}$.

In general, if the Hamiltonian has both positive and negative values, we may not use the Hilhorst,
the Prato nor the Lenzi integral transform formulae. To be able to apply integral transform
methods, we need to derive formulae that work under less restrictive conditions.

\section{Integral Transform Formulae for One-Dimensional Gases}

The derivation of more general formulae for one-dimensional gases is done here. Along the way, we will show an interesting feature of Tsallis statistical mechanics: for some values of $\beta$,
inaccessible regions of phase space may begin to appear. These inaccessible regions appear
because of the positive probability requirement.
The appearance of these inacessible regions in phase space is the reason for the failure of
integral transform methods for some values of $q$.
{\em Thus, previous workers who used the integral transform methods neglected, at some point in their
derivation, to
take the positive probability requirement into account.} This means their results need to be
reexamined.

In the following discussion, we concentrate on the canonical
ensemble. In particular, we use the unnormalized constraints formalism\cite{TsallisMendes}.
\subsection{The $q<1$ case}

The Tsallis partition function for the canonical ensemble is
\begin{eqnarray}
Z_q={\sum_i }'(1+\beta(q-1)H_i)^{\frac{1}{1-q}}\label{TsallisZ}
\end{eqnarray}

The prime in the summation means we have a {\em restricted} sum; we sum only over all states
for which $1+\beta(q-1)H_i \ge 0$. The $H_i$'s are the eigenvalues of the Hamiltonian.

It must be noted that in previous work, the restriction of positive
definite probabilities was not built into the notation. This ambiguity in the notation led
to conclusions which are difficult to justify physically because the positive probability requirement was forgotten.

Now suppose that we have a system of N identical particles in 1-dimension, with a Hamiltonian of the form
\begin{eqnarray}
H=\sum_{i=1}^N \frac{p_i^2}{2m}+U(x_1,...,x_N)\label{Hamiltonian}
\end{eqnarray}

Let us assume that the particles are confined in a box of length L. Then the partition function
becomes
\begin{eqnarray}
Z_q=\frac{1}{N!h^N}\int_0^{L}\mbox{d}x_1...\int_0^{L}\mbox{d}x_N\int\mbox{d}p_1...\mbox{d}p_N
[1+\beta(q-1)H]^{1/(1-q)}\label{Zustandsumme}
\end{eqnarray}
again, the integration is only over states for which $1+\beta(q-1)H \ge 0$. Now suppose that
the potential and $\beta$ have values within the box so that $1+\beta(q-1)U(x_1,...,x_N)>0$. Then
\begin{eqnarray}
&&\left[1-\beta(1-q)\sum_{i=1}^N \frac{p_i^2}{2m}+U(x_1,...,x_N)\right]^{1/(1-q)}\nonumber\\&&=[1+\beta(q-1)
U]^{1/(1-q)}
\left[1-\frac{\beta(1-q)\sum_{i=1}^N\frac{p_i^2}{2m}}{1-\beta(q-1)
U}\right]^{1/(1-q)}
\end{eqnarray}

We now use spherical coordinates in momentum space. We set $\sum_i^Np_i^2=P^2$, and note that the volume of an N-shell is $\mbox{d}V(P)=\frac{\pi^{N/2}NP^{N-1}}{(N/2)!}\mbox{d}P$.

We may introduce the momentum cut-off explicitly in the following manner: Let
\begin{eqnarray}
P_f(\beta)={\sqrt{\frac{2m[1-\beta(1-q)U]}{\beta(1-q)}}}.
\end{eqnarray}
From the positive probability requirement, we wish to sum over all states for which
$P \le P_f(\beta)$. Thus we may write
\begin{eqnarray} Z_q&=&\frac{1}{N!h^N}\int_0^{L}\mbox{d}x_1...\int_0^{L}\mbox{d}x_N\int
\mbox{d}p_1...\mbox{d}p_N[1+\beta(q-1)H]^{1/(1-q)}\nonumber\\
&=&\frac{1}{N!h^N}\int_0^{L}\mbox{d}x_1...\int_0^{L}\mbox{d}x_N(1-\beta(1-q)
U)^{1/(1-q)}\nonumber\\
& &\times \int_0^{P_f(\beta)} \mbox{d}P\frac{\pi^{N/2}NP^{N-1}}
{(N/2)!}
\left[1-\frac{P^2}{P_f(\beta)^2}\right]^{\frac{1}{1-q}}
\nonumber\\
&=&\frac{1}{N!h^N}\int_0^{L}\mbox{d}x_1...\int_0^{L}\mbox{d}x_N(1-\beta(1-q)
U)^{1/(1-q)}\nonumber\\
& &\times \frac{\pi^{N/2}N}{(N/2)!}\int_0^{P_f(\beta)}
P^{N-1}
\mbox{d}P
\left[1-\frac{P^2}{P_f(\beta)^2}\right]^{\frac{1}{1-q}}\label{Zq<1}
\end{eqnarray}

Note that the integral over the momenta can be evaluated using the Beta function\cite{Whittaker}
\begin{eqnarray}
B(p,q)&=&2\int_0^1x^{2p-1}(1-x^2)^{q-1}=\frac{\Gamma(p)\Gamma(q)}{\Gamma(p+q)}
\end{eqnarray}
where $\mbox{Re}(p)>-1$, and $\mbox{Re}(q)>-1$.

We then get, using the change of integration variables
$P=P_f(\beta)x$
\begin{eqnarray}
\int_0^{P_f}&&\mbox{d}P P^{N-1}
\left[1-\frac{P^2}{P_f(\beta)}\right]^{\frac{1}{1-q}}\nonumber\\
&=&\left[1-\beta(1-q)U\right]^{\frac{N}{2}}\left(\frac{2m}{\beta(1-q)}\right)^{\frac{N}{2}}
\int_0^1\mbox{d}x(1-x^2)^{\frac{1}{1-q}}x^{N-1}
\\
&=&2\left(\frac{2m}{\beta(1-q)}\right)^{N/2}[1-\beta(1-q)U]^{N/2} B\left(\frac{N}{2},1+\frac{1}{1-q}\right)\label{momInt}
\end{eqnarray}

Substituting the expression in (\ref{momInt}) into the expression for the partition function
(\ref{Zq<1}), we obtain
\begin{eqnarray}
Z_q&=&\frac{1}{N!h^N}\int_0^{L}\mbox{d}x_1...\int_0^{L}
\mbox{d}x_N(1-\beta(1-q)U(x_1,...,x_N))^{N/2+1/(1-q)}\frac{\pi^{N/2}N}{(N/2)!}\nonumber\\
& &\times 2\left(\frac{2m}{\beta(1-q)}\right)^{N/2}B\left(\frac{N}{2},1+\frac{1}{1-q}\right)
\label{intermediate}
\end{eqnarray}
Using the expression for the Beta function in terms of Gamma functions, and using the properties
of the Gamma function, we may simplify (\ref{intermediate})
\begin{eqnarray}
Z_q&=&\left(\frac{2m}{\beta(1-q)}\right)^{N/2}\frac{4\pi^{N/2}\Gamma(1/(1-q)+1)}{h^NN!
\Gamma((1/(1-q))+(N/2)+1)}\nonumber\\ & &\times \int_0^{L}\mbox{d}x_1...\int_0^{L}\mbox{d}x_N[1-\beta(1-q)U(x_1,...,x_N)]^{N/2+1/(1-q)}
\end{eqnarray}
Note that this expression is only valid when $1-\beta(1-q)U>0$ over the region of integration.
If we wish to consider ranges of $\beta$, or potentials $U$ for which this
restriction does not hold, we get the formula
\begin{eqnarray}
Z_q&=&\left(\frac{2m}{\beta(1-q)}\right)^{N/2}\frac{4\pi^{N/2}\Gamma(1/(1-q)+1)}{h^NN!
\Gamma((1/(1-q))+(N/2)+1)}\nonumber\\ & &\times \int_0^{L}\mbox{d}x_1...\int_0^{L}
\mbox{d}x_N[1-\beta(1-q)U(x_1,...,x_N)]^{N/2+1/(1-q)}\nonumber\\
& &\times\theta(1-\beta(1-q)U(x_1,...,x_N))\label{Zconf2Zq}
\end{eqnarray}
where the step function $\theta(x)$ is defined as
\begin{eqnarray}
\theta(x)&=&1 \quad \mbox{when} \quad x\ge 0\nonumber\\
&=&0 \quad \mbox{when} \quad x<0
\end{eqnarray}

We now introduce the Gamma function because we wish to write a power law form as an integral
of an exponential. The relevant representation of the Gamma function was used in the papers of Lenzi \cite{Lenzi},\cite{Fa}, and is given by \cite{Gradshteyn}:
\begin{eqnarray}
\frac{1}{\Gamma(z)}=\frac{\mbox{exp}(ab)b^{1-z}}{2\pi}\int_{-\infty}^{\infty}\frac{\mbox{e}^{itb}}
{(a+\mbox{i}t)^z}\mbox{d}t. \quad \mbox{where} \quad b>0\label{Gamma}
\end{eqnarray}

From (\ref{Gamma}), we have
\begin{eqnarray}
b^{z-1}=\frac{\Gamma(z)\mbox{exp}(ab)}{2\pi}\int_{-\infty}^{\infty}\frac{\mbox{e}^{itb}}
{(a+\mbox{i}t)^z}\mbox{d}t \quad \mbox{where} \quad b>0
\end{eqnarray}

Setting $a=1$, $b=1-\beta(1-q)U$, and $z=\frac{N}{2}+\frac{1}{1-q}+1$, we have
\begin{eqnarray}
[1-\beta(1-q)U]^{\frac{N}{2}+\frac{1}{1-q}}=\frac{\Gamma(\frac{N}{2}+\frac{1}{1-q}+1)}{2\pi}\int_{-\infty}^{\infty}\frac{\mbox{e}^{1+it}\mbox{e}^{-(1+it)\beta(1-q)U}}
{(1+\mbox{i}t)^z}\mbox{d}t\label{Power2}
\end{eqnarray}
Substituting (\ref{Power2}) into (\ref{Zconf2Zq}), we get
\begin{eqnarray}
Z_q&=&\left(\frac{2m}{\beta(1-q)}\right)^{N/2}\frac{4\pi^{N/2}\Gamma(1/(1-q)+1)}{h^NN!\Gamma((1/(1-q))+(N/2)+1)}\nonumber\\ & &\times \int_0^{L}\mbox{d}x_1...\int_0^{L}\mbox{d}x_N\frac{\Gamma(\frac{N}{2}+\frac{1}{1-q}+1)}{2\pi}\int_{-\infty}^{\infty}\frac{\mbox{e}^{1+it}\mbox{e}^{-(1+it)\beta(1-q)U}}
{(1+\mbox{i}t)^z}\mbox{d}t\nonumber\\
& &\times\theta(1-\beta(1-q)U(x_1,...,x_N))
\end{eqnarray}
Interchanging the order of integration, we get
\begin{eqnarray}
Z_q&=&\left(\frac{2m}{\beta(1-q)}\right)^{N/2}\frac{4\pi^{N/2}\Gamma(1/(1-q)+1)}{h^NN!2\pi}
\nonumber\\ & &\times \int_{-\infty}^{\infty}\mbox{d}t\frac{\mbox{e}^{1+it}}{(1+\mbox{i}t)^{\frac{N}{2}+\frac{1}{1-q}+1}}\nonumber\\&&\times\int_0^{L}\mbox{d}x_1...\int_0^{L}\mbox{d}x_N\mbox{e}^{-(1+it)\beta(1-q)U}
\theta(1-\beta(1-q)U)\label{q<1result}
\end{eqnarray}
The presence of the step function in (\ref{q<1result}) tells us that if $\beta$ is large enough,
then there ought to be inaccessible particle configurations.

Note that this expression may be further simplified if the Hamiltonian or the parameter $\beta$ is such that the condition $1-\beta(1-q)U>0$ is satisfied for every point in configuration space. If this condition is satisfied, then the expression for the $q<1$ partition function reduces to
\begin{eqnarray}
Z_q&=&\left(\frac{2m}{\beta(1-q)}\right)^{N/2}\frac{4\pi^{N/2}\Gamma(1/(1-q)+1)}{h^NN!2\pi}
\nonumber\\ & &\times \int_{-\infty}^{\infty}\mbox{d}t
\frac{\mbox{e}^{1+it}}{(1+\mbox{i}t)^{\frac{N}{2}+\frac{1}{1-q}+1}}
\int_0^{L}\mbox{d}x_1...\int_0^{L}\mbox{d}x_N\mbox{e}^{-(1+it)\beta(1-q)U}\\
&=&\left(\frac{2m}{\beta(1-q)}\right)^{N/2}\frac{4\pi^{N/2}\Gamma(1/(1-q)+1)}{h^NN!2\pi}
\nonumber\\ & &\times \int_{-\infty}^{\infty}\mbox{d}t
\frac{\mbox{e}^{1+it}}{(1+\mbox{i}t)^{\frac{N}{2}+\frac{1}{1-q}+1}}
Z_{conf}((1+it)\beta(1-q))\label{Zconf2Zq2}
\end{eqnarray}

Let us now discuss this very interesting result. First, the presence of the step function in
(\ref{q<1result}) tells us that if $\beta$ is large enough, then there ought to be
inaccessible particle configurations. That is, there exist points in phase space that
become inaccessible if we lower the temperature. This is in contrast to the usual
Gibbsian statistics, where lowering the temperature decreases, but does not totally
wipe out, the possibility of the system being in high energy states.

Second, when $1-\beta(1-q)U>0$ for every point in configuration space, then there is an
{\em exact} relationship between the Tsallis partition function and the Gibbs configuration
partition function. So, for potential energy functions that are bounded from above,
there is a temperature range for which (\ref{Zconf2Zq2}) is valid.

\subsection{The $q>1$ case}
We now consider the $q>1$ case. Again we pay careful attention to the restriction
of positive definite probabilities. The derivation is similar but not totally
identical to the derivation of the Hilhorst formula \cite{HilhorstFormula1}

Again, consider the partition function(\ref{Zustandsumme})
\begin{eqnarray}
Z_q=\frac{1}{N!h^N}\int_0^{L}\mbox{d}x_1...\int_0^{L}\mbox{d}x_N\int\mbox{d}p_1...\mbox{d}p_N
[1+\beta(q-1)H]^{1/(1-q)}
\end{eqnarray}

We must integrate over the region of phase space for which $1+\beta(q-1)H>0$. Again, we
introduce spherical coordinates so that we may write
\begin{eqnarray}
Z_q^{(1)}&=&\frac{1}{N!h^N}\int_0^{L}\mbox{d}x_1...\int_0^{L}\mbox{d}x_N(1-\beta(1-q)U(x_1,...,x_N))^{1/(1-q)}\nonumber\\
& &\times \int_{0}^{\infty}
\mbox{d}P\frac{\pi^{N/2}NP^{N-1}}
{(N/2)!}
\left[1-\frac{\beta(1-q)}{1-\beta(1-q)U}\frac{P^2}{2m}\right]^{\frac{1}{1-q}}\nonumber
\\ & &\times \theta\left(U-\frac{1}{\beta(1-q)}\right)\label{FirstPart}
\end{eqnarray}
and
\begin{eqnarray}
Z_q^{(2)}&=&\frac{1}{N!h^N}\int_0^{L}\mbox{d}x_1...\int_0^{L}\mbox{d}x_N(1-\beta(1-q)U(x_1,...,x_N))^{1/(1-q)}\nonumber\\
& &\times \int_{\sqrt{\frac{2m[1-\beta(1-q)U]}{\beta(1-q)}}}^{\infty}
\mbox{d}P\frac{\pi^{N/2}NP^{N-1}}
{(N/2)!}
\left[1-\frac{\beta(1-q)}{1-\beta(1-q)U}\frac{P^2}{2m}\right]^{\frac{1}{1-q}}\nonumber
\\ & &\times \theta\left(\frac{1}{\beta(1-q)}-U\right)\label{SecondPart}
\end{eqnarray}

The full partition function is given by
\begin{eqnarray}
Z_q=Z_q^{(1)}+Z_q^{(2)}\label{Zfull}
\end{eqnarray}

Let us now examine our results. The first term in the expression (\ref{Zfull}) for the full
partition function reduces to the Hilhorst formula when the Hamiltonian is positive
definite.

The second term represents the correction to the Hilhorst formula when the
Hamiltonian is not positive definite. Note that the presence of the step function tells us
that whenever $\left(\frac{1}{\beta(1-q)}-U\right)>0$ for a given point in configuration
space, then there is a lower limit to the allowed kinetic energies.

There is a problem with these expressions that we were unable to address: the convergence of the
integrals in expressions (\ref{FirstPart}) and(\ref{SecondPart}). If the integrals in
(\ref{FirstPart}) and in (\ref{SecondPart})do not converge for a given value of q, then there
might be reason to suppose that the probability distributions obtained are unnormalizable.

\section{Simple Applications}\label{1DGases}
\subsection{The Ideal Gas}
The ideal gas is the simplest classical gas.

We now reconsider the ideal gas within the unnormalized constraints formalism. Since
the Hilhorst formula is valid for the ideal gas in the $q>1$ regime and the since the $q>1$ partition function has already been worked out correctly \cite{HilhorstFormula1}, we work out the one-dimensional, $q<1$ case. Since the potential is always zero, we may use the results
of the previous section to immediately write the partition function as
\begin{eqnarray}
Z_q&=&\left(\frac{2m}{\beta(1-q)}\right)^{N/2}\frac{4\pi^{N/2}L^N\Gamma(1/(1-q)+1)}{h^NN!
\Gamma((1/(1-q))+(N/2)+1)}
\end{eqnarray}

On the other hand, Prato's formula\cite{Prato} leads to the result
\begin{eqnarray}
Z_q^{Prato}=\left(\frac{2m}{\beta(1-q)}\right)^{N/2}\frac{2\pi^{N/2}L^N\Gamma(1/(1-q)+1)}{h^NN!
\Gamma((1/(1-q))+(N/2)+1)}
\end{eqnarray}
Evidently, $Z_q^{Prato}<Z_q$. This is understandable because negative probability states are
included in the expression for the $q<1$ partition function.

\subsection{The Tonks gas}
The Tonks gas\cite{Lieb} is a one-dimensional gas with a hard-core potential energy function. If
the particles are a distance greater than $a$, then they act like free particles. However, they
cannot be nearer to each other than the distance $a$ because the potential energy becomes
infinite. In equations, the Tonks potential between two particles, (called the ith and jth
particles) is given by
\begin{eqnarray}
V_{Tonks}(x_i-x_j)&=& 0 \quad \mid x_i-x_j \mid \quad>a\nonumber\\
&=&\infty \quad \mid x_i-x_j \mid \quad\le a
\end{eqnarray}

It has been shown that the Tonks gas configuration partition function, in the Boltzmann-Gibbs treatment, is of the form
\begin{eqnarray}
Z_{conf}^{Tonks}=\frac{(L-(N-1)a)^N}{N!}.
\end{eqnarray}

Note that the Hamiltonian of the Tonks gas is positive definite. Thus, the Hilhorst formula
may be used to evaluate the $q>1$ partition function. Substituting the form of the Tonks gas
Boltzmann-gibbs partition function into the expression
\begin{eqnarray}
Z_q=\frac{1}{\Gamma(\frac{1}{q-1})}\int_0^{\infty}t^{\frac{2-q}{q-1}}\mbox{e}^{-t}
Z_1(t(q-1)\beta)\mbox{d}t,
\end{eqnarray}
the following is readily obtained
\begin{eqnarray}
Z_q^{Tonks}=\frac{(L-(N-1)a)^N}{N!\Gamma\left(\frac{1}{q-1}\right)}
\left(\frac{2m\pi}{\beta(q-1)}\right)^{N/2}
\Gamma\left(\frac{2-q}{q-1}\right)
\end{eqnarray}
\section{Conclusion}
The positive probability requirement (or the need for cut-off conditions) is a feature of Tsallis statistics that is not found in the usual statistics. It is this requirement that 
makes Tsallis statistics interesting and computationally nontrivial.

We chose to work within the unnormalised constraints formalism purely for mathematical convenience. There are, however, other Tsallis formalisms\cite{TsallisMendes} aside from the unnormalised constraints formalism. There is therefore a need for a reinvestigation of these other formalisms in the light of the positive probability requirement. 

We have shown, within the unnormalised constraints formalism, how integral transform methods may fail for some values of the parameter $q$. The failure of integral transform method emerges from the appearance of innaccessible phase space configurations; these innaccesible phase space configurations in turn arise from the positive probability requirement. 

We then worked out expressions for the $q<1$ and $q>1$ Tsallis partition function for one-dimensional gases made of identical particles that agree with the positive probability requirement. We were able to apply the formulae we obtained to the case of the Ideal Gas and 
the Tonks gas.  

\section*{Acknowledgments}
The authors would like to acknowledge the assistance of Mr. Augusto Morales,Jr.
\nocite{*}



\end{document}